\documentclass[aps,prd,showkeys,superscriptaddress,twocolumn,nofootinbib,floatfix]{revtex4-2}
\usepackage[applemac]{inputenc} 
\usepackage{amsmath,amssymb}
\usepackage{graphicx}
\usepackage{color}
\usepackage{hyperref}
\begin{document}
\title{New numerical framework for the generalized \\Baumgarte-Shapiro-Shibata-Nakamura formulation:\\ 
The vacuum case for spherical symmetry}
\author{M. A. Alcoforado}
\affiliation{Departamento de F\'{i}sica Te\'{o}rica, Universidade do Estado do Rio de Janeiro,
Rua S\~{a}o Francisco Xavier 524, 20550-013, Maracan\~{a}, Rio de Janeiro, RJ, Brazil}
\author{R. F. Aranha}
\affiliation{Departamento de F\'{i}sica Te\'{o}rica, Universidade do Estado do Rio de Janeiro,
Rua S\~{a}o Francisco Xavier 524, 20550-013, Maracan\~{a}, Rio de Janeiro, RJ, Brazil}
\author{W. O. Barreto}
\affiliation{Departamento de F\'{i}sica Te\'{o}rica, Universidade do Estado do Rio de Janeiro,
Rua S\~{a}o Francisco Xavier 524, 20550-013, Maracan\~{a}, Rio de Janeiro, RJ, Brazil}
\affiliation{Centro de F\'{i}sica Fundamental, Universidad de Los Andes, M\'{e}rida 5101, Venezuela}
\author{H. P. de Oliveira}
\affiliation{Departamento de F\'{i}sica Te\'{o}rica, Universidade do Estado do Rio de Janeiro,
Rua S\~{a}o Francisco Xavier 524, 20550-013, Maracan\~{a}, Rio de Janeiro, RJ, Brazil}

\begin{abstract}
Here we report a developed high performance and simplified version of the code denominated {\sc RIO}, which can be easily extended, for the generalized BSSN formulation. We implement a code which is regular at the center of symmetry, without use a special procedure for regularization, as usual.  We get exponential convergence for constraints. The numerical algorithm is based on the Galerkin-Collocation method developed successfully for diverse physical scenarios by the Numerical Relativity Group at {\sc UERJ}. For the sake of clarity in presentation, we consider here the most simple case to display the most salient features of the procedure. Thus, we focus on the definite tests of the new numerical framework. The timing and performance of the code show that we can reach a better accuracy close to the machine precision, for the Hamiltonian and momentum constraints.  {\sc RIO}  will be an open source code; currently is under continuous development to consider more general and realistic problems.    
\end{abstract}
\maketitle
\section{Introduction}
\par Numerical Relativity (NR) \cite{numrel} is nowadays one of the most solid areas in the context of Relativity, Gravitation and Astrophysics. Despite the lack of computational structure, its birth took place in the first half of the 1960's with the seminal work of Hahn and Lindquist \cite{hahn} and continued in the following decade with important articles by Smarr, Eppley and Piran \cite{smarr, eppley, piran}, where the first numerical evolutions and gravitational waves production were obtained. After this first moment, in approximately three decades, the computational capacity has increased significantly and large projects increased the scientific production as well. Among the main projects, we can highlight the {\it Binary Black Hole Grand Challenge Alliance} \cite{alliance} and the {\it Lazarus} project \cite{lazarus, lazarus2}. Despite these great efforts, the formalisms used for the construction of the algorithms had problems related to the hyperbolicity of the system of equations, which prevented their full integration. In 2005, Pretorius \cite{pretorius}, using harmonic coordinates, achieved a full integration of a collision of black holes including all phases of the coalescence. {This work together with \cite{campanellietal} and \cite{bakeretal} are considered a breakthrough in Gravitational Waves research.} 
This caused a considerable increase in the production of results using the various formalisms which take into account the strong hyperbolicity of the General Relativity equations. Nowadays, it is even possible to find computational consortia totally dedicated to the numerical integration of Einstein's equations \cite{etk}.
\indent \par From the point of view of the results, an entire effort to catalog the waveforms produced in the collision of compact objects was able to provide data to be compared with the observational data that would be obtained by the LIGO/Virgo consortium in the next years to come {\cite{ligo, virgo, boyleetal19,ninja1,ninja2,nrar,georgiatech,rit1,rit2,ncsa,sxs1}}. There is a very large community that takes care of the data analysis of these catalogs generated by the numerical relativity groups and that injects such data into the detection systems of the LIGO/Virgo consortium {\cite{nuria,LIGO-T1500606}. The detection of gravitational waves \cite{abbott1, abbott2,abbott3,abbott4,abbott5}} by astrophysical processes has further boosted the study and development of numerical relativity as a standard and powerful computational data source. 
\indent \par When considering the theoretical formalisms used, we can highlight the generalized harmonic (GH) \cite{gh}, the Baumgarte, Shapiro, Shibata and Nakamura (BSSN) \cite{bssn} and the Bona, Mass\'o, Seidel and Stela (BMSS) \cite {bmss} formalisms. These frames, although hyperbolic, are not covariant, which makes them impossible to use general curvilinear coordinate systems, that are interesting for certain physical systems. In a very well structured work, Brown \cite{b09} builds the 3D covariance of the BSSN equations, allowing several coordinate systems to be explored. Brown's formalism became known as generalized BSSN or G-BSSN in short. After this work, several authors presented algorithms for the solution of the G-BSSN equations in spherical coordinates \cite {ac15, bmcm13, am11}. A fundamental point for the realization of these codes is in the regularization of the system due to the singularities in the origin and in the polar axis. In this context, we can consider two types of regularization: i) with a fixed gauge \cite {bardeen, chop91}; and ii) with the introduction of new variables in place of the usual BSSN ones \cite {ag05}. In the case of the regularization scheme i), the fact of having a fixed gauge makes the gauge freedom unfeasible, which is one of the advantages of the BSSN formalism and which made, for instance, the ``maximal slice" ~and the ``1 + log" gauges so popular. In the case of ii), the choice of variables, apparently, is restricted to some type of symmetry such as the spherically symmetric and axisymmetric cases. An important work that circumvents these difficulties is that of Baumgarte et al. \cite{bmcm13}, where no type of symmetry is considered and no type of regularization is taken due to the use of a time evolution via the called partially implicit Runge-Kutta numerical method (PIRK) \cite{pirk}, in which two stages of integration are considered. In addition, the distribution of the points of the numerical grid avoids points on the origin and on the polar axis. Several tests with well-known systems were carried out and the convergence of the method proved to be quite satisfactory.
\par In this context, this work focuses on the construction of a solid code which deals with the regularization in a natural way for the solution of the G-BSSN equations in spherical coordinates. At first, we focus on a code that takes into account a spherically symmetrical system. We also chose to use a code based on the Galerkin-Collocation (GC) method, one of the main spectral methods available \cite{boyd}. {Spectral methods form a group of numerical methods as an alternative to finite difference and finite element methods. Within the context of numerical relativity were commonly used through {\it{Langage Objet pour la RElativit\'e Num\'eriquE}} (LORENE)  \cite{lorene} and {\it Spectral Einstein Code} (SpEC) \cite{spec}. Both projects contributed with excellent work (see for instance,  \cite{lorene_collaborations}, \cite{spec_collaborations})}

Over the years, the numerical relativity group at UERJ used spectral methods to solve several problems, mainly in the context of the characteristic formulation \cite {h01, h02}. Currently, the UERJ group is developing work aimed at building codes in the context of the 3+1 formalism for obtaining numerical initial data \cite{rio1} and, as can be seen in this work, developing our first code on the G-BSSN formalism via the Galerkin-Collocation method. Our purpose is to gather all the infrastructure built for these problems in a single repository with the name {\sc RIO} code \cite{rio2}. This is a way of establishing a method for storing and maintaining codes on a continuous and consistent way. 
\par Our GC code deals naturally with the regularization of the system by choosing a suitable and complete set of basis functions belonging to the Hilbert space $L^2[0,\infty]$. Such basis automatically satisfies all the necessary conditions for a regular behavior both at the origin and at the spatial infinity. The major advantage of spectral methods is in the exponential convergence of the solutions, as well as in the simplicity of implementation, once the basis is determined. Depending on the boundary conditions, each variable of the G-BSSN formalism has a minimally modified basis, but with the same collocation points which form the numerical grid.

{As a first application we have in mind the natural extension to consider gravitational collapse, with and without cosmological constant. This
includes a huge family of interesting problems, including holographic
ones. But our main goal currently is the extension to cylindrical 
coordinates to study 2D problems as Brill waves and rotating sources.}
\par That said, we have organized the content of the article as follows. In section II, we present the G-BSSN formalism with spherical symmetry and the choice of the initial data, as well as a discussion on the boundary conditions. In section III we introduce the Galerkin-Collocation method with a suitable choice of the basis functions which deals naturally with the regularization problem. The discretization of the computational grid in the radial coordinate is given through the collocation points associated with the chosen basis and, thus, we are able to reduce the system of partial differential equations in an autonomous dynamical system whose variables are given by the spectral coefficients of the approximate solution considered. Following this scheme, we build a fourth-order Runge-Kutta integrator (with a fixed step) for the system's time evolution. In section IV we present the numerical results, mainly with the convergence of the initial data as well of the Hamiltonian and momentum constraints. Here we also discuss the earlier results from the literature. Finally, in section V, we summarize the work as well as point out the future directions of our research. In this work $c=G=1$. 

\section{The Equations}
With the metric in the following form \cite{alcubierreetal}, \cite{ac15}
\begin{equation}
ds^2=-\alpha^2dt^2 +\psi^4(Adr^2 +Br^2d\Omega),
\end{equation}
where $\alpha$, $\psi$, $A$ and $B$ are functions of $t$ and $r$,
we use the G-BSSN formalism \cite{b09}, \cite{bmcm13}.  Thus, the evolution equations can be written as:
\begin{eqnarray}
\partial_t\alpha &=&-\alpha^2 K, \label{ee1}\\
\partial_t A &=& -2\alpha\tilde A_{rr},\label{ee2}\\
\partial_t B &=& -2\alpha\tilde A_{\theta\theta}, \label{ee3}\\
\partial_t \psi &=& -\frac{1}{6}\alpha\psi K, \label{ee4}\\
\partial_t\tilde\Lambda&=&\frac{2\alpha}{A}\left(\frac{6\tilde A_{\theta\theta}}{A}\frac{\partial_r\psi}{\psi} -\frac{2}{3}\partial_r K\right)+\nonumber\\
&&\frac{\alpha}{A}\left(\frac{ \tilde A_{rr}\partial_r A}{A^2} -\frac{2 \tilde A_{\theta\theta}\partial_r B}{B^2}  + \frac{4\tilde A_{\theta\theta}(A-B)}{rB^2}\right) - \nonumber \\
&& \frac{2 \tilde A_{rr}\partial_r\alpha}{A^2}, \label{ee5}\\
\partial_t\tilde A_{rr}&=& \frac{1}{\psi^4} \left[ -\mathcal{D}^{TF}_{rr}+\alpha R^{TF}_{rr}\right] +\nonumber\\
&& \alpha\left(\tilde A_{rr} K -\frac{2\tilde A_{rr}^2}{A}\right), \label{ee6}\\
\partial_t\tilde A_{\theta\theta}&=& \frac{1}{r^2\psi^4} \left[ -D^{TF}_{\theta\theta}+\alpha R^{TF}_{\theta\theta}\right] + \nonumber\\
&& \alpha\left(\tilde A_{\theta\theta} K -\frac{2\tilde A_{\theta\theta}^2}{B}\right), \label{ee7}\\
\partial_t K &=&\alpha\left(\frac{1}{3}K^2 +\frac{\tilde A_{rr}^2}{A^2}+\frac{2\tilde A_{\theta\theta}^2}{B^2}\right) -\mathcal{D}, \label{ee8}
\end{eqnarray}
{where $\tilde\Lambda$ is the radial component of the conformal connection $\tilde\Lambda^k$, 
$$\tilde A_{ij}=\text{diag}[\tilde A_{rr}(t,r), r^2\tilde A_{\theta\theta}(t,r),r^2\sin^2\theta \tilde A_{\theta\theta}(t,r)],$$ 
are the conformally rescaled trace-free part of the extrinsic curvature and $K$ is the trace of the extrinsic curvature (see \cite{ac15} for details).
Also follows that}
\begin{equation}
\mathcal{D}=\frac{1}{\psi^4}\left(\frac{\mathcal{D}_{rr}}{A}+\frac{2\mathcal{D}_{\theta\theta}}{r^2 B}\right),
\end{equation}

\begin{equation}
\mathcal{D}_{rr}=\partial^2_{r}\alpha -\frac{(\partial_r\alpha)}{2}\left(\frac{\partial_r A}{A}+\frac{{4}\partial_r\psi}{\psi}\right),
\end{equation}
\begin{equation}
\mathcal{D}_{\theta\theta}=\frac{r(\partial_r\alpha)}{A}\left[{B}+\frac{r}{2}\left({\partial_r B}+4B\frac{\partial_r \psi}{\psi}\right)\right],
\end{equation}

\begin{eqnarray}
R_{rr}&=&\frac{3(\partial_r A)^2}{4A^2}-\frac{(\partial_r B)^2}{2B^2}+A\partial_r \tilde\Lambda+\frac{1}{2} \tilde\Lambda \partial_r A+ \nonumber \\
&& \frac{1}{r}\left[-4\left(\frac{\partial_r\psi}{\psi}\right) -\frac{1}{B}(\partial_r A +2\partial_r B)+\frac{2A\partial_r B}{B^2}\right]-\nonumber \\
&&4\partial_r\left(\frac{\partial_r\psi}{\psi}\right) + 2\left(\frac{\partial_r\psi}{\psi}\right)\left(\frac{\partial_r A}{A}-\frac{\partial_r B}{B}\right)-\nonumber \\
&& \frac{\partial^2_{r} A}{2A}+ \frac{2(A-B)}{r^2 B},
\end{eqnarray}
\begin{eqnarray}
R_{\theta\theta}&=&\frac{r^2B}{A}\left[\frac{\partial_r\psi}{\psi}\frac{\partial_r A}{A}-2\partial_r\left(\frac{\partial_r\psi}{\psi}\right)-4\left(\frac{\partial_r\psi}{\psi}\right)^2 \right]+\nonumber \\
&&\frac{r^2}{A}\left[\frac{(\partial_r B)^2}{2B}-3\frac{\partial_r\psi}{\psi}\partial_r B-\frac{1}{2}\partial^2_{r} B +\frac{1}{2}\tilde\Lambda A \partial_r B\right] + \nonumber\\
&& r\left(\tilde\Lambda B-\frac{\partial_r B}{B}-6\frac{\partial_r\psi}{\psi}\frac{B}{A}\right) + \frac{B}{A} -1.
\end{eqnarray}
$\mathcal{D}^{TF}_{rr}$, $R^{TF}_{rr}$, $\mathcal{D}^{TF}_{\theta\theta}$ and $R^{TF}_{\theta\theta}$ are calculated
using 
\begin{equation}
{X}^{TF}_{rr}=\frac{2}{3}\left(X_{rr}-\frac{A X_{\theta\theta}}{Br^2}\right)
\end{equation}
\begin{equation}
{X}^{TF}_{\theta\theta}=\frac{1}{3}\left(X_{\theta\theta}-\frac{B X_{\theta\theta}}{A}\right)
\end{equation}
$X$ represents $\mathcal{D}$ or $R$, indistinctly.

And the Hamiltonian and the momentum constraints:

\begin{eqnarray}
\mathcal{H}&\equiv& \frac{2}{3}K^2 -\frac{\tilde A_{rr}^2}{A^2} - \frac{2\tilde A_{\theta\theta}^2}{B^2}+\nonumber\\
&&\frac{1}{\psi^4}\left(\partial_r\tilde\Lambda+\frac{1}{2}\tilde\Lambda\frac{\partial_r A}{A}+\tilde\Lambda\frac{\partial_r B}{B} +\frac{2\tilde\Lambda}{r}\right)-\nonumber\\
&&\frac{8}{A\psi^5}\left(\partial^2_{r}\psi-\frac{1}{2}\frac{(\partial_r A)\partial_r\psi}{A}+\frac{(\partial_r B)\partial_r\psi}{B}+\frac{2\partial_r\psi}{r}\right)-\nonumber\\
&&\frac{1}{A\psi^4}\left(\frac{1}{2}\frac{\partial^2_{r} A}{A}-	\frac{3}{4}\frac{(\partial_r A)^2}{A^2}+\frac{\partial^2_{r}B}{B}-\right. \nonumber\\
&&\left.\frac{1}{2}\frac{(\partial_r B)^2}{B^2}+\frac{2\partial_r B}{rB}+\frac{\partial_r A}{r B}\right) =0,\label{HC}\nonumber\\
\\
\mathcal{M}^r&\equiv&\frac{2}{3}\partial_r K-\frac{\partial_r \tilde A_{rr}}{A}-6\frac{\tilde A_{rr}}{A}\frac{\partial_r \psi}{\psi} +\frac{\tilde A_{rr}\partial_r A}{A^2}- \nonumber \\
&& \frac{(\partial_r B) \tilde A_{rr}}{AB}+\frac{(\partial_r B) \tilde A_{\theta\theta}}{B^2}-\frac{2\tilde A_{rr}}{rA}+\frac{2\tilde A_{\theta\theta}}{rB}=0. \label{MC}
\end{eqnarray}
{Observe that equations (\ref{ee1})-(\ref{ee8}) and (\ref{HC})-(\ref{MC}) are equivalents to the equations presented in \cite{am11} 
if we define 
\begin{equation}e^\chi=\psi, \,\, A_a=\tilde A_{rr}/A, \,\, A_b=\tilde A_{\theta\theta}/B, \,\, \hat\Delta=\tilde\Lambda, \label{defs_AM} \end{equation}
stressing that $\tilde A_{rr}/A + 2\tilde A_{\theta\theta}/B = 0$ ($\tilde A_{ij}$ is trace-free), and
\begin{equation}
\tilde\Lambda=\frac{1}{A}\left[\frac{\partial_r A}{2A} -\frac{\partial_r B}{B} - \frac{2}{r}\left(1-\frac{A}{B}\right)\right], \label{cc}
\end{equation}
which has to be used as a constraint because $\tilde\Lambda$ is considered in G-BSSN as an independent variable.} Also it is important to observe that we use the harmonic slicing given by Eq. (\ref{ee1}), with zero shift, instead the 1+log slicing $\partial_t\alpha=-2\alpha K$.
\subsection{Initialization}
{To evolve numerically the most simple case, pure gauge, we initialize the Minkowski spacetime by setting the conformal metric to $A(0,r)=B(0,r)=\psi=1$. The extrinsic curvature functions and the conformal connection are initialize to $K=\tilde A_{rr}=\tilde A_{\theta\theta}=\tilde\Lambda=0.$
The lapse function is set to 
\begin{equation}
\alpha(0,r)=1 + \frac{\alpha_{\iota} r^2}{1+r^2} 
\left[e^{-(r-r_0)^2/\sigma^2}+e^{-(r-r_0)^2/\sigma^2}\right], \label{ic}
\end{equation}
in order to compare our method and results with \cite{am11}. Clearly $\alpha_{\iota}$ represents an initial amplitude, $r_0$ the center of the Gaussian  and $\sigma$ its width.}

\subsection{Boundary conditions}
{The metric has to be conformally flat \cite{ac15} at the origin, 
that is, 
\begin{equation}
     A(t,0)=B(t,0).
\end{equation} 
Because we are using the Lagrangian choice we have
\begin{equation}
     AB^2=1,
\end{equation}
which leads to
\begin{equation}
A(t,0)=B(t,0)=1. \label{bc1}
\end{equation}
Also we can see that \cite{ac15}
\begin{equation}
\tilde A_{rr}(t,0)=\tilde A_{\theta\theta}(t,0)=0, \label{bc2}
\end{equation}
\begin{equation}
    \partial_r A(t,r)|_{r=0}=\partial_r B(t,r)|_{r=0}=0, \label{bc3}
\end{equation}
\begin{equation}
    \partial_r \tilde A_{rr}(t,r)|_{r=0}= \partial_r\tilde A_{\theta\theta}(t,r)|_{r=0}=0, \label{bc4}
\end{equation}
\begin{equation}
    \tilde\Lambda(t,0)=0, \label{bc5}
\end{equation}
\begin{equation}
    \partial_r K(t,r)|_{r=0}=\partial_r \psi(t,r)|_{r=0}=0. \label{bc6}
\end{equation}
Issues of parity are considered in the next section. 
At the outer boundary we set
\begin{equation}
A=B=\psi=\alpha=1, \,\,\text{at:}\,\, (t,\infty)
\end{equation}
and
\begin{equation}
\tilde A_{rr}=\tilde A_{\theta\theta}=K=\tilde\Lambda=0, \,\,\text{at:}\,\, (t,\infty).
\end{equation}}

According to definitions given by Eq. (\ref{defs_AM}) from now on we use $A_b=-A_a/2$, $A_a$, $\hat\Delta$ and $\chi$, stressing that the evolution equations are reduced to the number of seven without loss of generality. 
\begin{figure}[htb]
\includegraphics[height=7.0cm,width=7cm]{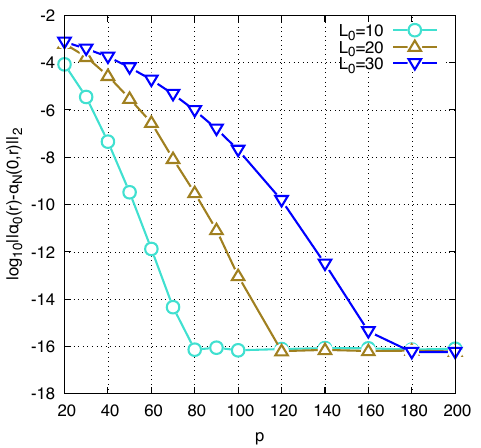}
\caption{Exponential convergence of  $||\alpha_0(r)-\alpha_N(0,r)||_2$ for the initial data given by Eq. (\ref{ic}) for different choices of $L_0$, with $\alpha_\iota=0.01$, $r_0=5$ and $\sigma=1$.}
\label{fig:ie}
\end{figure}
\begin{figure}[htb]
\includegraphics[height=4.5cm,width=7.5cm]{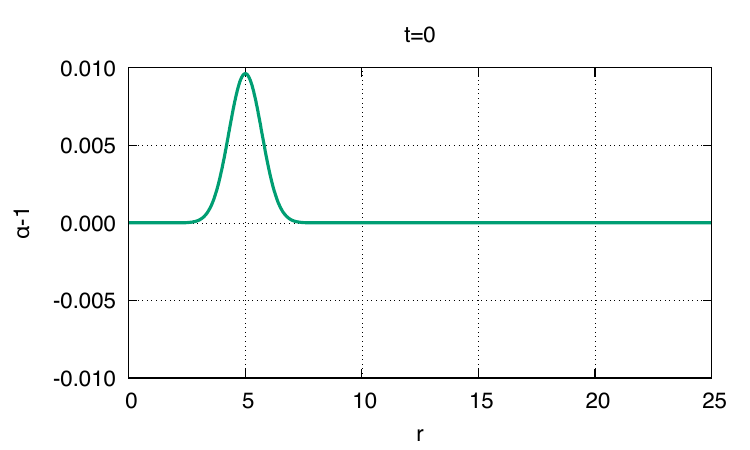}
\includegraphics[height=4.5cm,width=7.5cm]{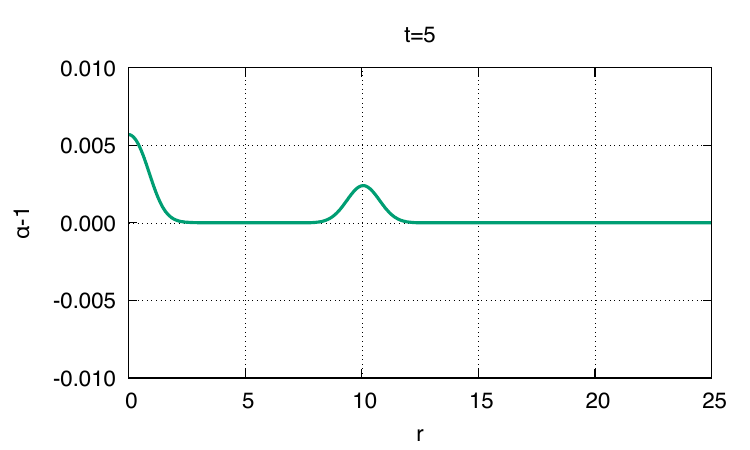}
\includegraphics[height=4.5cm,width=7.5cm]{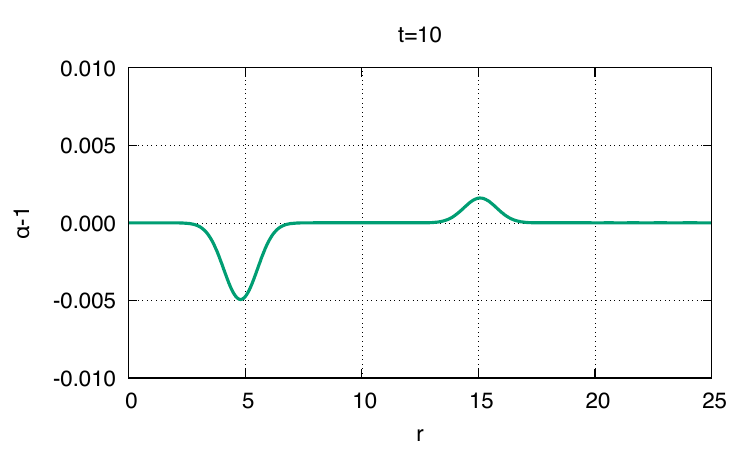}
\includegraphics[height=4.5cm,width=7.5cm]{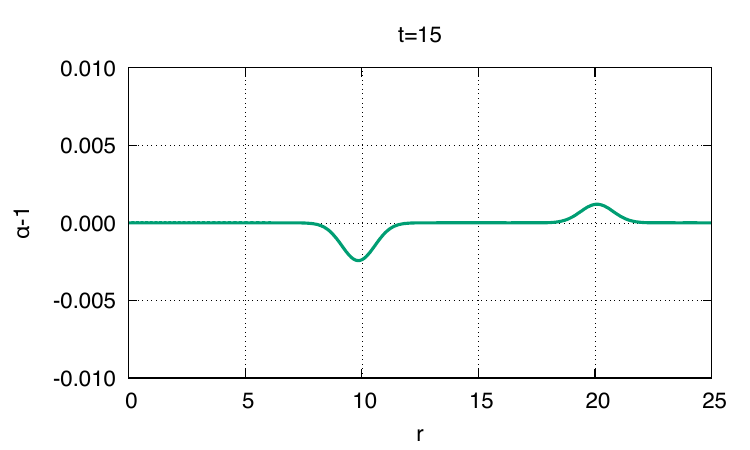}
\caption{Evolution of $\alpha-1$ for the initial condition given by Eq.(\ref{ic}) and parameters as given in Fig. \ref{fig:ie} with $p=320$, $L_0=30$ and $\Delta t=10^{-4}$.}
\label{fig:alpha}
\end{figure}
\begin{figure}[htb]
\includegraphics[height=5.5cm,width=8.5cm]{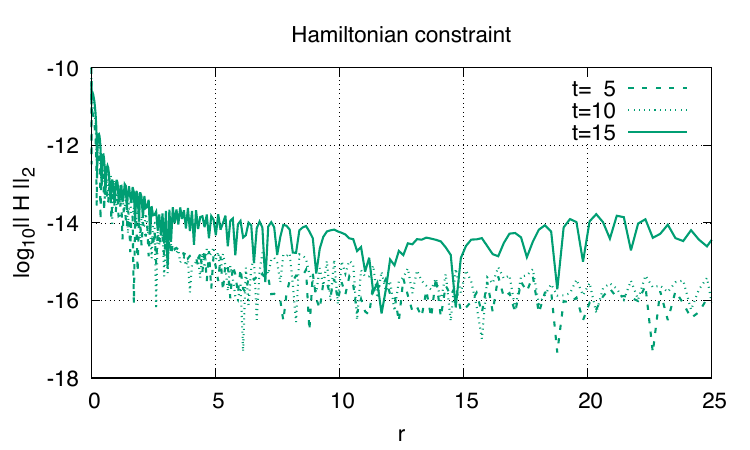}
\includegraphics[height=5.5cm,width=8.5cm]{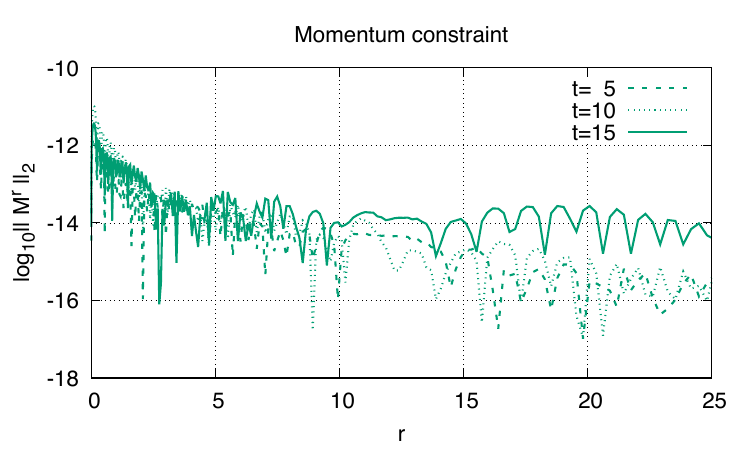}
\caption{Evolution of the Hamiltonian (upper panel) and momentum (lower panel) constraints, for the initial condition given by Eq.(\ref{ic}) and parameters as given in Fig. \ref{fig:ie} with $p=320$, $L_0=30$ and $\Delta t=10^{-4}$. The constraints are satisfied exactly at $t=0$.}
\label{fig:ce}
\end{figure}

\section{Numerical Method}
The use of any spectral method requires the correct choice of basis functions. In particular, for the Galerkin-Collocation method, we seek basis functions that automatically satisfy the boundary conditions. By dealing with the G-BSSN equations in spherical coordinates, there is another feature some functions must fulfill, namely, a definite parity with respect to expansion in $r$ as $r \rightarrow 0$. After inspecting the evolution equations (\ref{ee1}) - (\ref{ee8}), the functions $\alpha$, $K$, $\chi$, $A$, $B$ and $A_a$ have even parity, while ${\hat\Delta}$ has odd parity. For completeness, we have
\begin{eqnarray}
&&\alpha(t,r) = \alpha_0(t)+\alpha_2(t)r^2 + \mathcal{O}(r^4), \\
\nonumber \\
&&K(t,r) = K_0(t)+K_2(t)r^2 + \mathcal{O}(r^4), \\
\nonumber \\
&&\chi(t,r) = \chi_0(t)+\chi_2(t)r^2 + \mathcal{O}(r^4), \\
\nonumber \\
&&A(t,r) = 1 +\mathcal{O}(r^2), \\
\nonumber \\
&&B(t,r) = 1 +\mathcal{O}(r^2), \\
\nonumber \\
&&A_a(t,r) = \mathcal{O}(r^2).
\end{eqnarray}  

\noindent The above expressions satisfy the conditions (\ref{bc1})-(\ref{bc6}).
For the odd parity function $\hat{\Delta}$, we have
\begin{eqnarray}
\hat{\Delta}(t,r) = \hat{\Delta}_1(t)r + \mathcal{O}(r^3).
\end{eqnarray}  

According to Boyd \cite{boyd} there is just one class of basis functions derived from the standard Chebyshev polynomials with explicit parity for expansion of $r$ near the origin \textit{and} approach to zero as $r \rightarrow \infty$. These basis functions are the even and odd sines, $SB_{2n}(r)$ and $SB_{2n+1}(r)$, respectively. We can obtain these functions by defining 
\begin{eqnarray}
SB_0(r)=\left(1+\frac{r^2}{L_0^2}\right)^{-\frac{1}{2}}, \label{sb0}\\
\nonumber \\
SB_1(r)=\frac{2r}{L_0}\left(1+\frac{r^2}{L_0^2}\right)^{-1}, \label{sb1}
\end{eqnarray} 

\noindent where $L_0$ is the map parameter, and the recurrence formula
\begin{eqnarray}
SB_{n+1}(r)=\frac{2r}{L_0}\left(1+\frac{r^2}{L_0^2}\right)^{-\frac{1}{2}}SB_n(r)-SB_{n-1}(r), \label{sbn}\nonumber \\
\end{eqnarray} 

\noindent for all $n=1,2,..$.  Therefore, the functions $SB_{2n}(r)$ have  even parity with respect to $r=0$ and $SB_{2n+1}(r)$ odd parity {(see figure \ref{fig:basis_mapping} in the appendix).} 

At this point, we can establish the spectral approximations for each of the evolution variables:
\begin{eqnarray}
&\alpha_N(t,r)& = 1+\sum^{N}_{j=0}\,\hat{\alpha}_j(t) SB_{2j}(r), \label{bc1_}\\
\nonumber \\
&\chi_N(t,r)& = \sum_{j=0}^N\,\hat{\chi}_j(t) SB_{2j}(r), \label{bc2_}\\
\nonumber \\
&K_N(t,r)& = \sum_{j=0}^N\,\hat{K}_j(t) SB_{2j}(r), \label{bc3_}\\
\nonumber \\
&A_N(t,r)& = 1+\frac{1}{2}\sum_{j=0}^{N-1}\,\hat{A}_j(t) (SB_{2j+2}(r)-SB_{2j}(r)), \label{bc4_}\nonumber \\
 \\
&B_N(t,r)& = 1+\frac{1}{2}\sum_{j=0}^{N-1}\,\hat{B}_j(t) (SB_{2j+2}(r)-SB_{2j}(r)), \label{bc5_}\nonumber \\
 \\
&A_{aN}(t,r)& = \frac{1}{2}\sum_{j=0}^{N-1}\,\hat{A}_{aj}(t) (SB_{2j+2}(r)-SB_{2j}(r)), \label{bc6_}\\
\nonumber \\
&\Delta_N(t,r)& = \sum_{j=0}^{N-1}\,\hat{\Delta}_j(t) SB_{2j+1}(r), \label{bc7_}
\end{eqnarray}

\noindent In the above expressions, $\hat{\alpha}_j(t)$, $\hat{\chi}_j(t)$, $\hat{K}_j(t)$, $\hat{A}_j(t)$, $\hat{B}_j(t)$, $\hat{A}_{aj}(t)$, $\hat{\Delta}_j(t)$ are the modes or unknown coefficients and $N$ is the truncation order that limits the number of the modes. Notice the basis functions of $A(t,r)$, $B(t,r)$ and $A_a(t,r)$ are a combination of the even sines that behaves as $\mathcal{O}(r^2)$ near the origin providing the fullfilment of the conditions (\ref{bc1}) - (\ref{bc6}). We remark the combination of Chebyshev-like basis to satisfy the boundary conditions is one of the features of the Galerkin method.

We describe now the numerical algorithm based on the Galerkin-Collocation method \cite{h01,h02,rio1,rio2}. We first establish a convenient set of collocation points in the physical domain connected with the basis functions defined by Eq. (\ref{sb0}) - (\ref{sbn}). We use the following mapping \cite{boyd}:
\begin{eqnarray}
r_k = \frac{L_0 x_k}{\sqrt{1-x_k^2}}, \label{cpr}
\end{eqnarray}
where $$x_k=\cos\left(\frac{\pi k}{2N+2}\right), \,\, k=0,1,..,2N+2 \label{cpx}$$  
are the Chebyshev-Gauss-Lobatto points {(see figure \ref{fig:basis_mapping} in the appendix)}. The computational  domain $-1 \leq x \leq 1$ corresponds to $r \in (-\infty,\infty)$, but we are going to consider those points located in the region $0 \leq r < \infty$. In this case, we have the points $r_k$ for $k=1,2,..,N+1$, where $r_0 = \infty$ is excluded.

For the sake of illustration, we consider the evolution equation (\ref{ee1}) for the lapse function. The values of $\alpha(t,r)$ at the collocation points given by Eq. (\ref{cpr}), $\alpha_k(t)=\alpha_N(t,r_k)$, are related to the modes $\hat{\alpha}_j(t)$ through
\begin{eqnarray}
\alpha_k(t) = 1+\sum_{j=0}^N\,\hat{\alpha}_j(t)SB_{2j}(r_k),
\end{eqnarray} 

\noindent where $k=1,2,..,N+1$. In the code we write the above equation in the matrix form
\begin{eqnarray}
\begin{pmatrix}
\alpha_1\\
\alpha_2\\
\vdots\\
\alpha_{N+1}
\end{pmatrix}
=\begin{pmatrix}
1\\
1\\
\vdots\\
1
\end{pmatrix} + \mathbb{AL}\,
\begin{pmatrix}
\hat{\alpha}_1\\
\hat{\alpha}_2\\
\vdots\\
\hat{\alpha}_{N+1} \label{alpham}
\end{pmatrix},
\end{eqnarray}

\noindent where $\mathbb{AL}$ is the $(N+1) \times (N+1)$ matrix formed by the components  $\mathbb{AL}_{kj} = SB_{2j}(r_k)$ for $j=0,1,..,N$ and $k=1,2,..,N+1$. The lapse function can either be represented by their values at the physical space, $\alpha_k(t)$, or the spectral modes $\hat{\alpha}_j(t)$. 

We can obtain the residual equation associated with the evolution equation (\ref{ee1}). Taking into account the spectral approximations (\ref{bc1_}) and (\ref{bc3_}), we have

\begin{eqnarray}
\mathrm{Res}_\alpha(t,r) =  \partial_t \alpha_N + \alpha^2_N\,K_N.
\end{eqnarray}

\noindent The residual equation does not vanish due to the approximations for the lapse function and the extrinsic curvature. Following the prescription of the Collocation method, we assume that the residual equation vanishes at the collocation points \footnote{In the context of Method of Weighted Residuals \cite{finlayson}, it means that the test functions are the Dirac delta functions, $\delta(r-r_k)$.}. As a consequence, we obtain
\begin{eqnarray}
(\partial_t \alpha)_k = - \alpha^2_k K_k,\quad k=1,2,..,N+1,
\end{eqnarray}

\noindent where $K_k(t)$ are the values of the extrinsic curvature at the collocation points. There is a matrix equation that connects the values $K_k(t)$ and the modes $\hat{K}_j(t)$ similarly to Eq. (\ref{alpham}). The above set of $N+1$ first-order ordinary differential equations dictates the evolution of the values $\alpha_k(t)$.

We repeat the same procedure for each evolution equation with the caveat that Eqs. (\ref{ee2})--(\ref{ee8}) are automatically satisfied at the origin after taking into account the spectral approximations (\ref{bc1_}) - (\ref{bc7_}). Notice that from these approximations, the functions $A(t,r)$, $B(t,r)$, $A_a(t,r)$, and $\hat\Delta(t,r)$ are fixed at the origin, more specifically $A_N(t,0)=B_N(t,0)=1, A_{aN}(t,0)=\hat\Delta_N(t,0)=0$. Therefore, we have obtained the first-order sets with a total of $7N+3$ differential equations.  

There is a relevant remark. In the evolution equations for $K, A_a$, and $\hat\Delta$, we found several terms containing derivatives with respect to $r$. In order to calculate the values of these derivatives at the collocation, we use the corresponding modes. For instance, we express the values of $\partial_r \alpha$ in the following matrix form 
\begin{eqnarray}
\begin{pmatrix}
(\partial_r \alpha)_1\\
(\partial_r \alpha)_2\\
\vdots\\
(\partial_r\alpha)_{N+1}
\end{pmatrix}
=\mathbb{DAL}\,
\begin{pmatrix}
\hat{\alpha}_1\\
\hat{\alpha}_2\\
\vdots\\
\hat{\alpha}_{N+1}
\end{pmatrix},
\end{eqnarray}

\noindent where 
\[\mathbb{DAL}_{kj} = \left(\frac{\partial\,SB_{2j}}{\partial r}\right)_{r_k},\]

\noindent with $k=1,2,..,N+1$ and $j=0,1,..,N$. We have used the same procedure to calculate the values of all derivative terms such as $\partial_{r} \chi$, $\partial_{r} A$, $\partial_{r} B$, $\partial_{rr} \alpha$, and so on.

The evolution scheme proceeds as follows:

\begin{itemize}
	
\item  First, we have to provide the initial data 

\begin{eqnarray}
\alpha_0=\alpha(0,r), A_0=A(0,r), \nonumber \\
\nonumber \\
B_0=B(0,r),A_{a0}=A_a(0,r). \nonumber
\end{eqnarray}

\noindent The values $K(0)_k$ and $\chi(0)_k$, $k=1,2,..,N+1$ are obtained after solving the Momentum and Hamiltonian constraints, respectively. The initial values of $\hat\Delta_k(0),\,k=1,2,..,N$ arise from Eq. (\ref{cc}).

\item Next, from all values given at $t=0$, we can obtain all modes associated with the spectral approximations. In the sequence, we determine all values at the collocation points of the derivative terms present in the RHS of Eqs. (\ref{ee1})--(\ref{ee8}). 

\item With the evolution equations (\ref{ee1})--(\ref{ee8}), we calculate the initial values of 

\[(\partial_t \alpha)_k, (\partial_t K)_k, (\partial_t \chi)_k, \]

\noindent for $k=1,2,..,N+1$, and 

\[(\partial_t A)_k, (\partial_t B)_k, (\partial_t Aa)_k, (\partial_t \hat\Delta)_k, \]

\noindent for $k=1,2,..,N$.

\item Then, we determine the values $\alpha_k$, $K_k$,... at the next time step, and all the process repeats.

\end{itemize}

\noindent We have used a standard explicit fourth-order Runge-Kutta integrator to evolve the equations (\ref{ee1})--(\ref{ee8}).
\begin{figure}[htb]
\includegraphics[height=7.0cm,width=7.0cm]{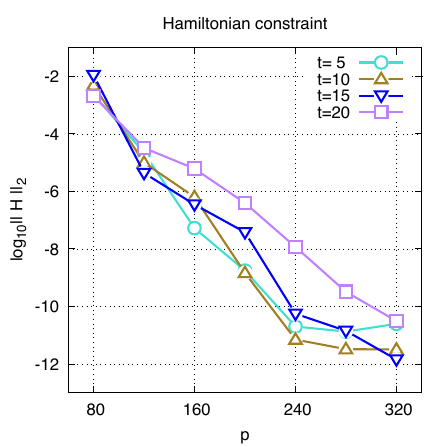}
\includegraphics[height=7.0cm,width=7.0cm]{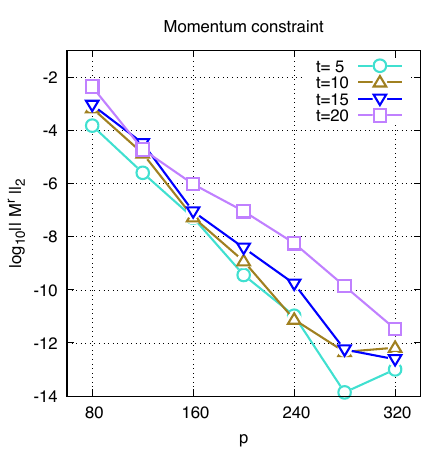}
\caption{Exponential convergence of the Hamiltonian (upper panel) and Momentum (lower panel) constraints with time, for the initial condition given by Eq.(\ref{ic}) and parameters as given in Fig. \ref{fig:ie} for $L_0=30$ and $\Delta t=10^{-4}$.}
\label{fig:cce}
\end{figure}
\section{Results}
In order to analyze our results we define an error measure by means of a Root Mean Square (RMS) norm $||...||_2$,
\begin{equation}
||...||_2=\left\{\frac{1}{2}\int_0^{\infty}(...)^2 dr \right\}^{1/2},
\label{eq:L2}
\end{equation}
where $(...)$ is an expected computational zero. We take the initial condition given by Eq. (\ref{ic}) and we calculate the spectral modes to numerically reconstruct the initial data $\alpha_N(0,r)$; thus $(...)=\alpha_0(r)-\alpha_N(0,r)$ in this case. We perform the numerical integration using a Legendre-Gauss-Lobatto quadrature off the collocation points. The figure  \ref{fig:ie} shows the expected exponential convergence for this initial data. Note that the convergence depends on the parameter $L_0$, but in any case the convergence can be exponential.  If we observe the mapping of the collocation points (Fig. \ref{fig:basis_mapping} in the appendix), it is clear that the Gaussian pulse at $t=0$ with peak at $r_0=5$ is better resolved if we use $L_0=10$, instead $L_0=20$ or $L_0=30$. This fact is clearly displayed in Fig. \ref{fig:ie} in terms of the rate of convergence. {We have to resolve traveling pulses with the best choice of the map parameter}. {The bad choice of $L_0$ may leads to evolutions with an exponential increasing (until saturation) of the error in the constraints.} {We have a caveat for this behavior. When the pulses inevitably enter a region with wide spacing between the collocation points, the error associated to both constraints increases. With the present code, this problem is solved simply by increasing the truncation order and the map parameter $L_0$. We stress that for a fixed truncation order we can still increase $L_0$.}
 {For instance, considering an integration until $t=20$, after tests and calibration the selection $L_0=30$ and $p=320$ for the initial condition (\ref{ic}) leads to a clear exponential convergence. On the other hand, if we select $L_0=10$ and $p=320$, we observe an increase that could be exponential (until saturation) of the Hamiltonian and momentum constraints.} {For a long time evolution, even with the appropriate choice of $L_0$ and $p$, may arise spurious oscillations or artifacts. In that case the multi-domain decomposition and dissipation techniques could be useful to ameliorate these issues.}
 We evolve the system of equations for the G-BSSN, using a standard fourth order Runge-Kutta method, with a fixed time step of $\Delta t= 10^{-4}$. Particularly we display in figure \ref{fig:alpha} the evolution of the lapse function $\alpha$. The evolution goes as is expected. The initial data contains two components; one begins to travel to the left and the other to the right. The left component hits the center and is reflected (freely at $r=0$) to travel then inverted to the right. Meanwhile the other initial component travels to the right. The maximum peak is separated from the minimum of the reflected part (travelling to the right) in ten spatial units, to keep this separation while the trend goes away. Clearly the peaks (maximum and minimum) travel with speed one, as expected, to the left and to the right, in any stage of the evolution. Observe that the advance in time corresponds to the same displacement in space. We stress here that we do not impose any special condition at the origin, neither a special numerical treatment. We could deal with non linear evolutions, which we hope the {\sc RIO} code is able to solve as well. We focus here on the definite tests of a new numerical framework, using the Galerkin-Collocation method. Figure \ref{fig:ce} shows the evolution of the constraints (\ref{HC}) and (\ref{MC}) for the same evolution displayed in Fig. \ref{fig:alpha}. {Initially the constraints are satisfied exactly, but it is clear that the constraints loss accuracy with evolution for any choice of parameters. For our choice after calibration the error in both constraints is under control for the considered interval of time, being maximum at $r=0$ of order $\sim 10^{-11}$ and decreasing to $\sim 10^{-14}-10^{-16}$ in the displayed region of evolution.}
To analyze this behavior carefully we study the convergence in time for the constraints. Now $(...)$ in (\ref{eq:L2}) is represented by the constraint equations (\ref{HC}) and (\ref{MC}), respectively. Thus, Fig. \ref{fig:cce} displays the exponential convergence of the constraints. Clearly the accuracy shown in Fig. \ref{fig:ce} is consistent with the accuracy observed in Fig. \ref{fig:cce} for $p=320$, at the monitored times. What it is most important, the convergence of the constraints improves exponentially with the truncation order. {For the evolved initial data in this work (up to $t=20$) with $p=320$, the accuracy of the Hamiltonian constraint is between  $10^{-11}$ and $10^{-12}$, and for the momentum constraint between $10^{-11}$ and $10^{-14}$.}  Our results are in complete accord with Ref. \cite{am11}, but our accuracy is largely better without any additional cost for regularization at $r=0$. {As a matter of fact, still we can evolve for a larger $p$ up to reach saturation ($\sim 10^{-16}$) in the constraint errors.} 

At this point we briefly consider some development, platform, timing and performance issues. We develop our prototype code with Maple (version 18), also we prepare partial versions for Octave and Python. But the main complete and structured development used to obtain the results presented here was in Fortran (from scratch, with open source libraries and compiler). A serial Fortran Rio code (modular and remarkable simple) was calibrated with the Maple prototype, up to some point. The final Fortran code is not dependent of the Maple prototype, running {\it ab initio} specifying only a set of parameters for the G-BSSN problem. We ran the Fortran code on an Intel core i7-9700k@8x4.9 GHz with 64 Gb of memory, under Ubuntu 18.04 bionic. For a fixed $p$ the Fortran code scales linearly with the maximal time of the evolution. For example, for $t_{max}=15$ used for figures \ref{fig:alpha} and \ref{fig:ce} the timing was $\approx 10\,$ min. For $p=320$, the maximum truncation used here the code runs overnight (eight hours); the required memory is negligible even for the largest used truncation. By far we can run to improve the accuracy, but in the present case is not necessary.
The RIO code is under continuous development to deal with more general relativistic problems; it will be an open source code.  

\begin{figure}[htb]
\includegraphics[height=7.cm,width=7.cm]{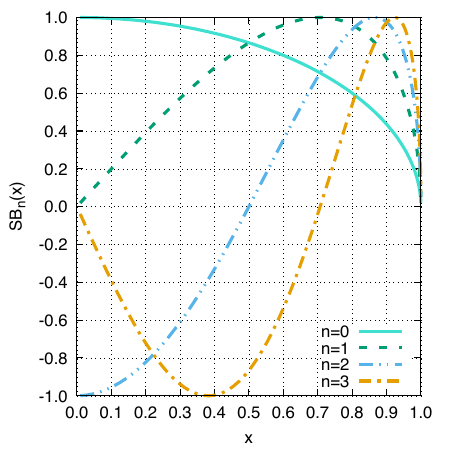}
\includegraphics[height=6.8cm,width=6.8cm]{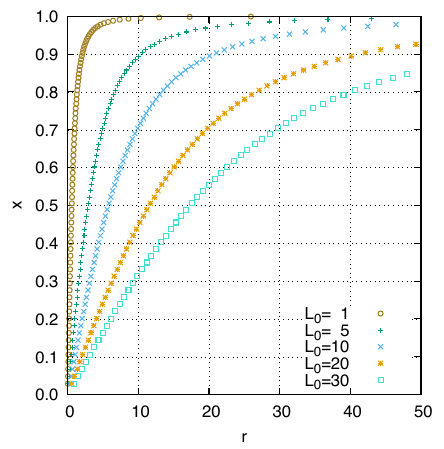}
\caption{First four basis-functions given by Eqs. (\ref{sb0})-(\ref{sbn}) (upper panel) and distribution of the collocation points for $p=80$ and different  $L_0$, using Eq. (\ref{cpr}) (lower panel).}
\label{fig:basis_mapping}
\end{figure}

\section{Conclusions}
In this paper we developed a RIO code \cite{rio2} version for the G-BSSN. We use spherical coordinates and consider the most simple case, evolving the pure gauge. The remarkable simplified code is exponentially convergent, as expected. The regularization at the origin is not a computational problem; using the Galerkin-Collocation method we avoid any complication or special treatment there. As a matter of fact, the basis choice for each field guarantee a regular
behavior close to and at the boundaries. {We will apply our code to problems in which 
the constraint errors do not run away owing the lack of spatial resolution within the evolved time. The calibration for each specific problem will be necessary.}
We will extend the treatment to include matter fields and/or other spatial dimensions. {We have in mind gravitational collapse, cosmological scenarios like post-inflationary pre-heating models, and holographic simulations for heavy ions collisions.} 
{The critical collapse is an excellent problem to test any code, even under spherical symmetry,
with a black hole formation near $r = 0$, requiring a map parameter $L_0 \le 1$, typically.
In this respect, the evolution time to study the critical phenomena is very short in comparison with the evolution 
time considered in the present work.}

{We have implemented the BSSN formulation using {\it spherical coordinates}, which are
suitable the study gravitational collapse in 1D/2D/3D. As have been studied by other authors (see Ref. \cite{bmcm13}), the extension 
to higher dimensions is natural and straightforward within the G-BSSN. In this sense we are currently considering the implementation 
of the G-BSSN as proposed by Brown (Ref. \cite{b09}) for the particular choice of {\it cylindrical coordinates}. 
As far as we know there is not such implementation reported in the literature. We expect to apply this last version of the {\sc RIO} code to the implosion of Brill waves \cite{hwb17}, \cite{lk21}.} 

Here we only display the most salient features of a new and simple tool to deal with more complex problems elsewhere, particularly with nonlinear evolutions. Some work is in progress considering cylindrical coordinates. 

\begin{acknowledgments}
 W.B. acknowledges the financial support by Universidade do Estado do Rio de Janeiro (UERJ) and Funda\c{c}\~{a}o Carlos Chagas Filho de Amparo \`{a} Pesquisa do Estado do Rio de Janeiro (FAPERJ); also thanks the hospitality of the Departamento de F\'\i sica Te\'orica at Instituto de F\'\i sica, UERJ. H.P.O. acknowledges the financial support of Brazilian Agency CNPq. 
\end{acknowledgments}

\appendix
\section*{Appendix: Basis and mapping}
For sake of completitude we show in figure \ref{fig:basis_mapping} the first four basis-functions given by Eqs. (\ref{sb0})-(\ref{sbn}) and how the collocation points are distributed by the parameter $L_0$, using Eq. (\ref{cpr}). 

\end{document}